\documentclass[twocolumn,english,aps,pre,showpacs,superscriptaddress,floats,amsmath,amssymb,floatfix]{revtex4-1}

\usepackage[T1]{fontenc}
\usepackage[utf8]{inputenc}
\usepackage{color}
\usepackage{amsmath}
\usepackage{graphicx}
\usepackage{graphics}
\usepackage{gensymb}
\usepackage{amssymb}
\usepackage{acronym}
\usepackage{mathtools}
\usepackage{bm}

\makeatletter
\@ifundefined{textcolor}{}
{
 \definecolor{BLACK}{gray}{0}
 \definecolor{WHITE}{gray}{1}
 \definecolor{RED}{rgb}{1,0,0}
 \definecolor{GREEN}{rgb}{0,1,0}
 \definecolor{GREEN2}{rgb}{0,0.4,0}
 \definecolor{BLUE}{rgb}{0,0,1}
 \definecolor{CYAN}{cmyk}{1,0,0,0}
 \definecolor{MAGENTA}{cmyk}{0,1,0,0}
 \definecolor{YELLOW}{cmyk}{0,0,1,0}
 \definecolor{YELLOW2}{cmyk}{0,0,1,0.6}
 \definecolor{ORANGE}{rgb}{1,0.22,0}

 }
\makeatother

\def\vec#1{\mathbf{#1}}
\def\vF{\vec{F}}
\def\vG{\vec{G}}
\def\va{\vec{a}}
\def\vf{\vec{f}}

\def\vr{\vec{r}}
\def\vs{\vec{s}}
\def\vv{\vec{v}}
\def\vx{\vec{x}}
\def\vw{\vec{w}}
\def\vz{\vec{z}}
\def\vpsi{\boldsymbol{\psi}}
\def\vxi{\boldsymbol{\xi}}

\begin{document}
\typeout{\the\linewidth}

\title{The Inertial Spin Model of flocking with position-dependent forces}

\author{Sebastián Carruitero}
\affiliation{Instituto de F\'isica de L\'iquidos y Sistemas Biol\'ogicos (IFLySiB), CONICET and Universidad Nacional de La Plata, Calle 59 n.\ 789, B1900BTE La Plata, Argentina}
\affiliation{CCT CONICET La Plata, Consejo Nacional de Investigaciones Cient\'\i{}ficas y T\'ecnicas, Argentina}
\affiliation{Departamento de F\'\i{}sica, Facultad de Ciencias Exactas, Universidad Nacional de La Plata, Argentina}

\author{Alejo Costa Duran}
\affiliation{Instituto de F\'isica de L\'iquidos y Sistemas Biol\'ogicos (IFLySiB), CONICET and Universidad Nacional de La Plata, Calle 59 n.\ 789, B1900BTE La Plata, Argentina}
\affiliation{CCT CONICET La Plata, Consejo Nacional de Investigaciones Cient\'\i{}ficas y T\'ecnicas, Argentina}
\affiliation{Departamento de F\'\i{}sica, Facultad de Ciencias Exactas, Universidad Nacional de La Plata, Argentina}

\author{Giulia Pisegna}

\affiliation{Department of Living Matter Physics, Max Planck Institute for Dynamics and Self-Organization, D-37077 G\"ottingen, Germany}

\author{Mauricio B. Sturla}
\affiliation{Instituto de F\'isica de L\'iquidos y Sistemas Biol\'ogicos (IFLySiB), CONICET and Universidad Nacional de La Plata, Calle 59 n.\ 789, B1900BTE La Plata, Argentina}
\affiliation{CCT CONICET La Plata, Consejo Nacional de Investigaciones Cient\'\i{}ficas y T\'ecnicas, Argentina}

\author{Tomás S. Grigera}
\affiliation{Instituto de F\'isica de L\'iquidos y Sistemas Biol\'ogicos (IFLySiB), CONICET and Universidad Nacional de La Plata, Calle 59 n.\ 789, B1900BTE La Plata, Argentina}
\affiliation{CCT CONICET La Plata, Consejo Nacional de Investigaciones Cient\'\i{}ficas y T\'ecnicas, Argentina}
\affiliation{Departamento de F\'\i{}sica, Facultad de Ciencias Exactas, Universidad Nacional de La Plata, Argentina}
\affiliation{Istituto dei Sistemi Complessi, Consiglio Nazionale delle Ricerche, Rome, Italy}

\begin{abstract}

  We propose an extension to the \acf{ISM} of flocking and swarming.  The model has been introduced to explain certain dynamic features of swarming (second sound, a lower than expected dynamic critical exponent) while preserving the mechanism for onset of order provided by the Vicsek model.  The \ac{ISM} has only been formulated with an imitation (``ferromagnetic'') interaction between velocities.  Here we show how to add position-dependent forces in the model, which allows to consider effects such as cohesion, excluded volume, confinement and perturbation with external position-dependent field, and thus study this model without periodic boundary conditions.  We study numerically a single particle with an harmonic confining field and compare it to a Brownian harmonic oscillator and to a harmonically confined active Browinian particle, finding qualitatively different behavior in the three cases.

\end{abstract}

\maketitle

\acrodef{ISM}{inertial spin model}

\section{Introduction}
\label{sec:intro}

Collective animal motion \cite{sumpter_principles_2006, Vicsek2012} is a particularly striking aspect of emergent collective behavior where 
collective order (as in flocks of birds flying together) can arise from simple short-range interactions between individuals.  Several models have been proposed to describe flocking behavior, dating back at least to the 1980s \cite{aoki1982, reynolds1987}.  The paradigmatic Vicsek model of flocking \cite{Vicsek1995a, ginelli_physics_2015} (and the related Toner-Tu field theory \cite{toner_long-range_1995, toner_flocks_1998, toner2012}) has received much attention from the statistical physics community because it predicts the appearance of ordered flocks starting with a simple set of microscopic local rules, similar to the way ferromagnetic order arises in the Ising or Heisenberg models.  It turns out that the phase diagram of the Vicsek model is more complicated than that of the classic ferromagnetic order-disorder transition, featuring a discontinuous transition and region with microphase separation (see \cite{chate2020} for a review); however it remains true that it allows to explain how a flock with fully ordered velocities can result from a local ``ferromagnetic'' imitation rule.

Although the Vicsek model is successful in explaining many aspects  of flocking at the static, or stationary, level, it is not suitable to interpret certain dynamical features related to the presence of inertial effects.  The finding of wave-like propagation of direction information during turns in starling flocks \cite{attanasi_information_2014} led to the proposal of the \acf{ISM} \cite{Cavagna2014}, which we consider here.  The model is described in detail in Sec.~\ref{sec:original-ism}, but it is essentially the Vicsek model endowed with a Hamiltonian-like (second order) dynamics, so that second time derivative of the velocity is proportional to the effective social force, instead of the first time derivative as in Vicsek.  Although it is true that in the thermodynamic limit the large-scale behavior is described by an over-damped theory \cite{cavagna2019a}, namely the Toner-Tu \cite{toner_flocks_1998} hydrodynamic theory (which is a coarse-grained version of the Vicsek model), inertial effects can be observed in finite systems \cite{cavagna_silent_2015}.  This is relevant in the description of observations of biological flocks, which at sizes of a few thousands of individuals are large but still far enough from the thermodynamic limit that finite-size effects are important.  In the language of the renormalization group (RG), a crossover phenomenon arises in the RG flow such that at intermediate sizes the global properties will be described by an inertial fixed point with an unstable direction, instead of the stable over-damped fixed point that rules in the thermodynamic limit \cite{cavagna2019}.

The dynamic critical behavior of midge swarms \cite{cavagna_dynamic_2017} also displays inertial effects.  At moderate system size, the Vicsek transition looks continuous \cite{Chate2008}, and it can be used to interpret static aspects of swarms such as the presence of scale-free correlations \cite{Attanasi2014}, but it is again insufficient to account for the dynamic behavior.  In particular the dynamic critical exponent of the Toner-Tu theory \cite{chen2015} is higher than the experimental value.  A coarse-grained version of the \ac{ISM} was recently employed \cite{cavagna2023a} to show that both inertia and activity are needed to explain the observations.

The \ac{ISM} was introduced in refs.~\cite{attanasi_information_2014, Cavagna2014} (see also the review \cite{cavagna_physics_2018}).  The formation of flocks at $T=0$ was considered in \cite{ha2019} and \cite{markou2021}, while the finite-temperature equilibrium in the mean-field case was studied in \cite{benedetto2020} and \cite{ko2023}, and \cite{huh2022} considered a variant incorporating uniform external fields controlling alignment and rotation, but  with a sort of mean-field spin.  In these works, as well as in the numerical simulations of \cite{Cavagna2014, cavagna2023a}, infinite space or periodic boundary conditions were employed.

In this work we consider extending the \ac{ISM} to include position-dependent forces.  In the original \ac{ISM} the interaction is between velocities (actually, velocity directions), such that particles tend to align the velocities with each other, just as in the Vicsek model.  Positions enter the picture only indirectly, through the definition of the interaction network (which can be metric or topological \cite{kumar2021}).  There are several reasons why positional forces are desirable.  To treat finite systems with metric interactions it is necessary to introduce some kind of confinement, because otherwise small velocity fluctuations eventually lead to particles becoming isolated from the flock and thus ``evaporating'' the system to infinite dilution if the boundary conditions are open.  This can be avoided using periodic boundary conditions, reflecting boundaries, external fields, or inter-particle interactions providing cohesion.  Reflecting walls for confinement \cite{chepizhko2021} or partial confinement \cite{tu1998, morin2018, moreno2020}, and cohesion through attractive forces \cite{chate2008a} have been considered for the Vicsek case.  Position-dependent inter-particle interactions may also be included for other purposes, e.g.\ to add an excluded volume potential that will limit the local density \cite{tu1998}.  Furthermore, an effective spatial confinement mechanism, mediated by external landmarks, is crucial for natural swarms to maintain global cohesion. Incorporating this element into our theoretical model will thus offer a more comprehensive understanding of the biological system. This will help clarify the system's bulk properties in the presence of confinement as well as the significance of boundaries in the swarm formation. A position-dependent external field can also be interesting to study perturbation and response: a field coupled to the velocities is straightforward to add (as it has been done for Vicsek \cite{kyriakopoulos_leading_2016}), but experimentally it may be easier to impose a perturbation with a position-dependent field, for example by implementing a moving artificial marker to perturb an insect swarm \cite{Attanasi2014a}.

Given the Hamiltonian structure of the ISM, which involves velocity and spin as canonical variables, it is not immediately obvious what is the best way to add a position-dependent force.  We discuss this in the next section.  After proposing a consistent way to implement these forces, we explore numerically some results regarding field-induced confinement.

\section{Inertial Spin model with external position-dependent forces}

\begin{figure}
  \includegraphics[width=\columnwidth]{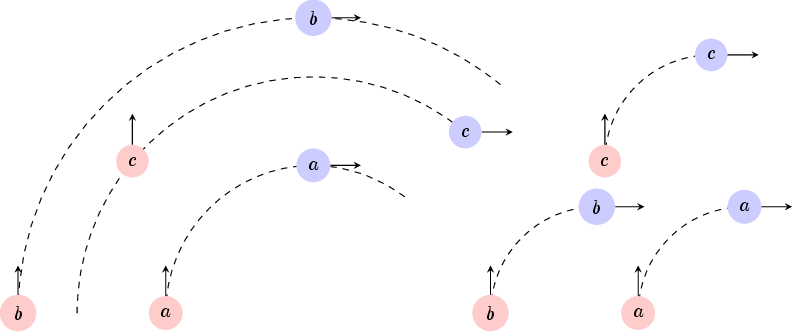}
  \caption{Parallel path vs.\ equal radius turns.  Two 3-particle flocks are shown making a clockwise 90 degree turn (light red circles represent the initial positions of the flock, light blue circles the final positions).  In the parallel path case (left), the result is a rigid body rotation around a common center; this rotation is generated by usual angular momentum.  Instead, in the equal radius case (right), which is required if all particles move at the same speed, the final configuration does not result from a rigid body rotation: note that the particle at the front of the flock is $c$ before the turn and $a$ afterwards.  This transformation is generated by the spin, which rotates particles' velocities.}
  \label{fig:turns}
\end{figure}

Originally, the \ac{ISM} was proposed \cite{attanasi_information_2014, Cavagna2014} based on the experimental observation of second sound (i.e.\ order-parameter waves) in the turning of starling flocks \cite{attanasi_information_2014}.  Second sound is an indication that an inertial mechanism is at work that results in propagation with constant speed (vs.\ diffusive propagation as it occurs in the overdamped case).  The goal is then to formulate a model with the static properties of Vicsek's model (i.e.\ capable of spontaneously producing orientational order) but with inertial rather than diffusive dynamics \cite{cavagna_physics_2018}.  It seems thus reasonable to seek a Hamiltonian formulation, as this will lead naturally to canonical (i.e.\ inertial) equations of motion.  One then needs to identify the correct canonical coordinate / conjugate momentum pair.  A key observation is that when a flock changes direction of motion individual birds turn following paths of approximately the same radius, rather than following parallel paths (see Fig.~\ref{fig:turns}).  This suggests that invariance under \emph{internal} (rather than global) rotations is the relevant symmetry.  Indeed, equal-radius turns are not generated by rigid rotations (as are parallel-path turns), but by rotations of the internal orientation of each individual, which moves at approximately constant speed.  These considerations lead to recognize the Vicsek interaction as an interaction between \emph{orientations,} rather than velocities, and to propose a Hamiltonian formalism in which the particle orientation is the canonical coordinate.  Its canonical conjugate, the \emph{spin,} is the generator of \emph{internal} rotations.

In a Hamiltonian theory, the symmetry under internal rotations leads to conservation of the spin (the corresponding momentum).  Indeed this quantity can be conserved in an equal-radius turn, in contrast to the angular momentum (generator of global, rigid rotations).  Angular momentum could be conserved in a turn following parallel paths, but this is forbidden by the requirement of constant speed.  However, spin conservation cannot be expected to hold exactly, so dissipation of the spin is introduced by adding noise and friction terms (as in the standard Langevin equation) in the spin equation of motion.  This is how temperature enters the theory, making it possible to tune the system between order and disorder.

\subsection{The original ISM}
\label{sec:original-ism}

Let us first introduce the original \ac{ISM} and then our proposal to
include position-dependent forces.  For what follows, it is convenient
to derive the \ac{ISM} using the velocity of the $i$-th particle
$\vv_i$ and its canonical conjugate, instead of following the original
presentation \cite{attanasi_information_2014, Cavagna2014,
  cavagna_physics_2018} employing orientation and spin.  We do not
consider speed fluctuations, so that a set of hard constraints
\begin{equation}
  g_i(\vv_i) = \vv_i^2 - v_0^2 = 0
\end{equation}
will be imposed.  Note however that $\vv_i$ is treated as an internal \emph{canonical coordinate,} and completely unrelated, as far as the Hamiltonian formalism is concerned, to the usual mechanical momentum.  Its canonical momentum, which we call $\vw_i$, is defined by
\begin{equation}
   \{v_i^\mu,w_j^\nu\} = \delta_{ij} \delta^{\mu\nu},
\end{equation}
where $\{\ldots\}$ are Poisson brackets.  The Hamiltonian formalism applies to the space of the internal degrees of freedom $(\vv_i,\vw_i)$; the connection between $\vv_i$ and the actual particle velocity is made through an extra equation
\begin{equation}
    \dot{\vr_i} = \vv_i,
\end{equation}
which complements the canonical equations of motion.  Since $\vr_i$ is also a parameter of the Hamiltonian (via the potential), the whole theory is in this sense pseudo-Hamiltonian.

One proposes a Hamiltonian
\begin{equation}
  \mathcal{H} = \sum_i \frac{w_i^2}{2\mu} + \mathcal{V}(\{\vv_i\}) + \sum_i\lambda_i g_i(\vv_i). \label{eq:oISM-H-w}
\end{equation}
The first term, which has the form of a kinetic energy, introduces inertia, with $\mu$ a social, or effective, mass.   The role of the third term is to enforce the constraints through the Lagrange multipliers $\lambda_i$.  The second term is an interaction potential, which in the \ac{ISM} is chosen to implement Vicsek's \cite{Vicsek1995a} velocity-imitation.  In continuous time this is obtained from
\begin{equation}
  \mathcal{V}(\{\vv_i\}) = \frac{J}{v_0^2} \sum_{ij} n_{ij} \vv_i \cdot \vv_j,
\end{equation}
with $J$ a coupling constant and $n_{ij}$ the adjacency matrix, which defines the interaction network ($n_{ij}=1$ if $i$ and $j$ are interacting neighbours, and 0 if not). 
Positions enter into $\mathcal{H}$ through the adjacency matrix, which can be defined to implement metric (through a cut-off radius) or topological (by choosing a fixed number of nearest neighbors) interactions.  As far as the Hamiltonian formalism is concerned, positions are just parameters of the potential, but the extra equation $\dot{\vr}_i=\vv_i$ turns them into coordinates, and the model into an active model.

The canonical equations of motion are
\begin{subequations}
\begin{align}
  \dot{\vv}_i &= \frac{\partial \mathcal{H}}{\partial \vw_i} = \frac{\vw_i}{\mu}, \label{eq:oISMp-v} \\
  \dot{\vw}_i &= -\frac{\partial \mathcal{H}}{\partial \vv_i} = \vf_i - 2\lambda_i\vv_i, \label{eq:oISMp-w}
\end{align}
\label{eq:oISMp-vw}
\end{subequations}
where $\vf_i = \partial \mathcal{V}/\partial \vv_i$ is the interaction force.  The Lagrange multipliers can be found using the relations $\dot{\vv}_i\cdot\vv_i=0$ and $\left(\dot{\vv}_i\right)^2 + \vv_i\cdot\ddot{\vv}_i=0$, which follow from the first and second time derivatives of the constraint.  Eq.~\eqref{eq:oISMp-w} is then
\begin{equation}
  \dot{\vw}_i = \vf_i^\perp - \frac{\left(\dot{\vv}_i\right)^2}{v_0^2} \vv_i,
\end{equation}
with $\perp$ denoting the projection in the direction perpendicular to the velocity:
\begin{equation}
  \va^\perp = \va - (\va\cdot\vv_i) \frac{\vv_i}{v_0^2} = -\frac{1}{v_0^2} \vv_i \times \left(\vv_i\times \va\right).
\end{equation}

Now it is convenient to introduce the spin of the $i$-th particle as $\vs_i = \vv_i\times\vw_i$.  The Poisson brackets
\begin{equation}
  \label{eq:spin-def}
  \{s_i^\mu,v_j^\nu\} = \delta_{ij} \epsilon^{\mu\nu\rho} v^\rho, \qquad
  \{s_i^\mu,w_j^\nu\} = \delta_{ij} \epsilon^{\mu\nu\rho} w^\rho,
\end{equation}
where greek superindices indicate cartesian components, $\delta_{ij}$ is Kroneker's delta and $\epsilon^{\mu\nu\rho}$ is the completely antisymmetric tensor, show that $\vs_i$ does generate the internal rotations.  Using the constraint again, the condition $\vv_i\cdot\vw_i=0$ allows to express the momentum in terms of the spin, $\vw_i = - \vv_i \times\vs_i / v_0^2$ and the Hamiltonian can be written as
\begin{equation}
    \mathcal{H} = \sum_{i} \frac{\mathbf{s}_i^2}{2\chi} + \mathcal{V}(\{\vv_i\})    \label{eq:HISM-orig},
\end{equation}
defining $\chi = v_0^2\mu$.  The final equations of the \ac{ISM} are
\begin{subequations}
  \begin{align}
    \dot{\vr}_i & = \vv_i, \label{eq:rdot-nofield}\\
    \dot{\vv}_i &= \frac{1}{\chi}\vs_i \times \vv_i , \label{eq:vdot-nofield}\\
    \dot{\mathbf{s}}_i &= \mathbf{v}_i\times \left[\frac{J}{v_0^2} \sum_j n_{ij}\mathbf{v}_j-\frac{\eta}{v_0^2}\dot{\mathbf{v}}_i+\frac{1}{v_0}\boldsymbol{\xi}_i \right],\label{eq:sdot-nofield}
  \end{align}
  \label{eq:explicit-nofield}
\end{subequations}
where we have added the friciton and stochastic terms, with $\langle \xi_i^\mu(t) \rangle =0$, $\langle \xi_i^\mu(t) \xi_j^\nu(t') \rangle = 2 \eta T \delta_{ij}\delta^{\mu\nu} \delta(t-t')$.  Note that~\eqref{eq:vdot-nofield} and the deterministic part of~\eqref{eq:sdot-nofield} can be derived either substituting $\vs_i$ for $\vw_i$ in \eqref{eq:oISMp-vw} or directly from the Hamiltonian \eqref{eq:HISM-orig} and the Poisson brackets
\begin{subequations}
 \begin{align}
    \dot{\mathbf{v}}_i &= \left\{ \mathbf{v}_i, \mathcal{H} \right\} = - \mathbf{v}_i \times \frac{\partial \mathcal{H}}{\partial \mathbf{s}_i} ,\\
    \dot{\mathbf{s}}_i &= \left\{ \mathbf{s}_i, \mathcal{H} \right\} = - \mathbf{v}_i \times \frac{\partial \mathcal{H}}{\partial \mathbf{v}_i} - \vs_i\times \frac{\partial \mathcal{H}}{\partial \vs_i}.
  \end{align}
\end{subequations}
In contrast, \eqref{eq:rdot-nofield} does not follow from $\{\vr,\mathcal{H}\}$.

Let us stress again that the Hamiltonian structure of the ISM involves $\mathbf{v}_i$ and $\mathbf{w}_i$ (or $\mathbf{s}_i$), and not the positions.  For example, in 2-$d$ one can show that the \emph{velocity orientations} interact like in an inertial Kuramoto model \cite{markou2021}.  Adding an external force that acts directly on the velocities is thus as straightforward as adding an external field in a Kuramoto model: one can add a term $\mathbf{h}\cdot\sum_i\mathbf{v}_i$ in the Hamiltonian, and this fits in perfectly within the Hamiltonian equations, because $\mathbf{v}_i$ is a canonical variable.  On the other hand, $\mathbf{r}_i$  is \emph{not} a canonical variable, but the integral of $\mathbf{v}_i$, i.e.\ something analogous to the integral of the Kuramoto orientations, which is a non-trivial object to treat within the Hamiltonian formalism.  For this reason, adding a force that depends on $\vr_i$ is not straightforward.  This is what we discuss next.

\subsection{Position-dependent forces}
\label{sec:intr-posit-depend}

Now we want to introduce position-dependent forces and fields.  While a velocity-dependent field can be added quite naturally in the formalism as a new term $H_\textrm{field}(\mathbf{v}_i)$ in Hamiltonian \eqref{eq:HISM-orig}, position-dependent forces require some thought because position is not part of the canonical variables of this Hamiltonian.  There are in principle two ways to add these forces.  One can expect that they should appear in \eqref{eq:vdot-nofield}, or~\eqref{eq:oISMp-v}, just as in ordinary Newton's equation.  Alternatively, one can argue that the forces should appear in \eqref{eq:sdot-nofield} or~\eqref{eq:oISMp-w}, since it is this equation that encodes the inertial mechanism that controls motion in this model.  We shall add both kind of force, but it will turn out that both can be handled as a new force in the spin equation, provided it depends on the velocity in a specific way.

The forces must be added in a way that respects the constraint of constant speed, so that it is convenient to start with \eqref{eq:oISM-H-w}, which uses $\vw_i$ and includes the constraint explicitly through the multipliers $\lambda_i$.  We propose to add the forces writing
\begin{equation}
  \begin{split}
  \label{eq:hamilt-with-w-and-lagmult}
  \mathcal{H} = &{} \sum_i\frac{w^2}{2\mu} + \mathcal{V}(\{\vv_i\}) - \sum_i \vv_i\cdot \vF(\{\vr\}) \\ & + \sum_i \vw_i\cdot \frac{\vG(\{\vr\})}{m} + \sum_i\lambda_i (v_i^2 - v_0^2),
  \end{split}
\end{equation}
where $\vG$ and $\vF$ are the new position-dependent forces and we have allowed the possibility that the mass associated to the positional force, $m$, is different from $\mu$.  The corresponding canonical equations are
\begin{subequations}
  \begin{align}
    \dot{\vv}_i &= \frac{\partial \mathcal{H}}{\partial \vw_i} = \frac{\vw_i}{\mu} + \frac{\vG_i}{m}, \label{eq:ISMnew-p-v} \\
    \dot{\vw}_i &= -\frac{\partial\mathcal{H}}{\partial \vv_i} = -\frac{\partial \mathcal{V}}{\partial \vv_i} + \vF_i - 2\lambda_i\vv_i. \label{eq:ISM-intermediate}
  \end{align}
\end{subequations}

So far our choice of the new terms in the Hamiltonian has been motivated by how the equation of motion will look, i.e.\ whether the force will act directly on the acceleration $\dot{\vv}_i$, or indirectly through $\dot{\vw}_i$.  However, looking directly at the Hamiltonian one may wonder whether more general terms, coupling non-linearly to $\vv_i$ or $\vw_i$, would be better choices.  The coupling to $\vv_i$ is standard in Hamiltonian systems, since $\vv_i$ plays the equivalent of a position, and can certainly be made non-linear. However, this is clearly not what we want here, because that would lead to forces dependent on both $\vv_i$ and $\vr_i$, and we wish to separate the interactions between velocities (the term $\mathcal{V}(\{\vv_i\})$) and those dependent on positions.  As for the term proportional to $\vw_i$, it turns out that it would be problematic to choose it non-linear in $\vw_i$.  One needs that the terms involving $\vw_i$ (the ``kinetic energy'') be convex, so that the Hamiltonian can be related to a corresponding Lagrangian via a Legendre transformation.  A non-convex Lagrangian leads to multi-valued Hamiltonian, and ill-defined dynamics \cite{chi2014}.  Although non-convex Lagrangians have been discussed for quantum systems and to study spontaneous breaking of time-translation invariance \cite{henneaux1987, shapere2012, chi2014}, we here we'd rather stick to the classical Hamiltonian formalism.  Non-linear couplings can be introduced at the price of restricting the parameters (e.g.\ the coefficients of a polynomial in $\vw_i$) so that convexity is preserved (as for example in \cite{casiulis2020a}, where a Hamiltonian with terms both linear and quadratic in the velocities is considered), we avoid this because it could lead to imposing restrictions on the form of $\vG(\{\vr_i\})$.   

We thus restrict ourselves to linear couplings to $\vv_i$ and $\vw_i$, and proceed to eliminate the $\lambda_i$ from the equations of motion using the fact that first and second time derivatives of the constraints must vanish to find
\begin{equation}
  \lambda_i = \frac{\mu}{2v_0^2} \left( \frac{\vw_i}{\mu} + \frac{\vG_i}{m} \right)^2 + \frac{\vv_i}{2v_0^2} \cdot \left(\vf_i +\vF_i \right) + \frac{\mu}{2m v_0^2} \dot{\vG}_i \cdot\vv_i.
\end{equation}
Eq.~\eqref{eq:ISM-intermediate} becomes then
\begin{equation}
  \dot{\vw}_i = \vf_i^\perp + \vF_i^\perp -\frac{\mu}{v_0^2}\left( \frac{\vw_i}{\mu} + \frac{\vG_i}{m}\right)^2 \vv_i - \frac{\mu}{m v_0^2} (\dot{\vG}_i \cdot \vv_i) \vv_i.
  \label{eq:ISMintermedate2}
\end{equation}

As before, the equations assume a simpler form if one introduces the spin.  Defining $\vs_i=\vv_i\times\vw_i$, this can be inverted with the help of the constraint to find $\vw_i = -\mu (\vv_i\cdot\vG_i) / (m v_0^2) - \vv_i\times \vs_i / v_0^2$ and rewrite \eqref{eq:ISMnew-p-v} and \eqref{eq:ISMintermedate2} as
\begin{subequations}
  \label{eq:eom-sv-notfinal}
  \begin{align}
    \dot{\vv}_i &= -\frac{1}{\chi} \vv_i\times \left[ \vs_i + \frac{\mu}{m} \vv_i\times \vG_i\right], \label{eq:new-with-old-spin} \\
    \dot{\vs}_i &= \vv_i\times \left(\vf_i+\vF_i\right) - \frac{\chi}{X^2} (\vv_i\cdot\vG_i) (\vG_i\times \vv_i) \nonumber \\
                &  - \frac{1}{X} \vG_i\times (\vv_i \times \vs_i),
  \end{align}
\end{subequations}
with $\chi=v_0^2 \mu$, $X=v_0^2 m$, which reduce to~\eqref{eq:explicit-nofield} (apart from noise) when $\vF_i = \vG_i=0$. Eqs.~\eqref{eq:eom-sv-notfinal} can be further simplified, because not every form of $\vG_i$ will lead to observable changes in particle trajectories.  The situation is similar to the case of the Hamiltonian formulation of electromagnetic forces, where a gauge transformation alters the vector potential but not the trajectories.  This can be seen by redefining the momentum and spin:  Eqs.~\eqref{eq:ISMnew-p-v} and~\eqref{eq:new-with-old-spin} suggest to define $\vz_i = \vw_i + (\mu/m) \vG_i$, and  $\tilde \vs_i = \vv_i\times\vz_i$.  Renaming $\tilde \vs_i \longrightarrow \vs_i$ and adding the stochastic terms the equations of motion finally read
\begin{subequations}
  \label{eq:ISM-field-final}
  \begin{align}
    \dot{\vv}_i &= \frac{1}{\chi} \vs_i\times\vv_i, \\
    \dot{\vs}_i &= \vv_i\times \left[ -\frac{\partial \mathcal{V}}{\partial \vv_i} + \vF_i + \frac{\chi}{X}(\vv_i\cdot\nabla) \vG_i \right] + \notag\\
    & \vv_i \times \left[-\frac{\eta}{v_0^2} \dot{\vv}_i + \frac{1}{v_0}\vxi_i \right].
  \end{align}
\end{subequations}

The final equations of motion cannot be written from Poisson brackets, because $\vr_i$ is not a canonical coordinate, and in consequence $\dot{\vG}_i = (\vv_i\cdot\nabla) \vG_i$ does not follow naturally from $\{\vG,\mathcal{H}\}$.  However it is possible to write a slightly different Hamiltonian that will yield~\eqref{eq:ISM-field-final} from $\dot{\vv}_i = \{\vv_i,\mathcal{H}\}$, $\dot{\vs}_i = \{\vs_i,\mathcal{H}\}$, provided $\vG_i$ can be written as a gradient, $\vG_i=\nabla \Gamma_i$:
\begin{equation}
  \begin{split}
    \mathcal{H} ={}& \sum_i\frac{s_i^2}{2\chi} + \mathcal{V}(\{\vv_i\}) - \sum_i \vv_i\cdot\vF_i(\vr_i) + \\
    &\frac{\mu}{2m} \sum_{i\mu\nu} v_i^\mu v_i^\nu \frac{\partial^2  \Gamma(\vr_i)}{\partial r_i^\mu \partial r_i^\mu}.
  \end{split}
\end{equation}

\subsection{Overdamped limit and dimensionless quantities}
\label{sec:overdamped-limit}

It is possible to eliminate $\vs_i$ and write a single, second-order, equation of motion for the velocity.  This can be more convenient for numerical integration, as then one can apply one of the well-known discretizations used in molecular or stochastic dynamics \cite[see e.g.][]{cavagna_spatio-temporal_2016}:
\begin{multline}
  \chi \frac{\ddot{\vv}_i}{v_0} + \chi \frac{(\dot{\vv}_i)^2}{v_0^2} \frac{\vv_i}{v_0} + \eta \frac{\dot{\vv}_i}{v_0} = \\J \sum_j n_{ij}\frac{\vv_j^\perp}{v_0} + v_0 \vF^\perp +  \frac{\chi}{X} \left(\frac{\vv_i}{v_0}\cdot\nabla\right) v_0^2\vG_i^\perp  + \vxi^{\perp}
  \label{eq:second-order}
\end{multline}
Writing the equation in this way makes it clear that the lhs is independent of $v_0$ (in effect, we are writing an equation of motion for the orientation),  and so must the rhs.  The interaction force is clearly independent of $v_0$ (as it should since it comes from a potential energy involving orientations), but in this form we realize that the positional forces must be such that $v_0\vF_i$ and $v_0^2\vG_i$ are independent of the speed.  For what follows it's then convenient to introduce $\vpsi_i=\vv_i / v_0$.

We now seek the overdamped limit of~\eqref{eq:second-order}.  Rescaling the time $t\to \hat t= t/a$ we have $\hat\vx_i(\hat t) = \hat\vx_i(t/a)=\vx_i(t)$, $\hat \vv_i(\hat t) = d\hat\vx / d\hat t = a \vv$, $\hat v_0=a v_0$, $\hat \vpsi(\hat t) = \hat\vv/\hat v_0 = \vpsi(t)$, and $\dot{\hat\psi} = a \dot\psi$, so that
\begin{multline}
  \frac{\chi}{a^2} \ddot{\hat\vpsi}_i + \frac{\chi}{a^2} (\dot{\hat\vpsi}_i)^2 \hat\vpsi_i + \frac{\eta}{a} \dot{\hat\vpsi}_i = \\
J \sum_j n_{ij} \hat\vpsi_j^\perp + v_0 \vF^\perp +  \frac{\chi}{X} \left(\hat\vpsi_i\cdot\nabla\right) v_0^2\vG_i^\perp  +
  \sqrt{\frac{2T\eta}{a}}\vxi^{\perp}(\hat{t}),
\end{multline}
where we have redefined $\vxi$ so that $\langle \vxi_i(\hat t) \vxi_j({\hat t}') \rangle = \delta(\hat t - \hat t' )\delta_{ij}$.  Choosing $a\propto \eta$ one arrives at an equation that will become first order in the limit $\chi/\eta^2\to0$.  It is convenient that $a$ have units of time so that $\hat t$ is non-dimensional.  We choose $a=\eta/J$, and write the equations of motion as
\begin{subequations}
  \begin{align}
    \dot {\hat \vx} & = \hat v_0 \hat \vpsi(\hat t) \label{eq:final-nondim-x} \\
    \Omega \ddot{\hat\vpsi}_i + \Omega (\dot{\hat\vpsi}_i)^2 \hat\vpsi_i + \dot{\hat\vpsi}_i & =  \sum_j n_{ij} \hat\vpsi_j^\perp + \hat\vF^\perp  \notag\\
             & \quad +  \frac{\chi}{X} \left(\hat\vpsi_i\cdot\nabla\right) \hat\vG_i^\perp  \notag\\
             & \quad + \sqrt{\frac{2T}{J}} \vxi^{\perp}(\hat{t}),
  \end{align}
\end{subequations}
where we have defined
\begin{subequations}
  \begin{align}
    \Omega &= \frac{J\chi}{\eta^2}, \label{eq:defOmega} \\
    \hat \vF_i & = \frac{v_0}{J} \vF_i, \\
    \hat \vG_i & = \frac{v_0^2}{J} \vG_i.
  \end{align}
\end{subequations}
All these quantities are dimensionless (provided one chooses the units of $\vG$ so that $\mu$ and $m$ have the same units so that $\chi/X$ is non-dimensional, see Eq.~\eqref{eq:ISMnew-p-v}).

$\Omega$ is a measure of the relative weight of inertia vs.\ dissipation.  The overdamped limit is obtained taking $\Omega\to0$, in which case the equation for the orientation becomes first-order, yielding a continuous-time version of the Vicsek model:
\begin{subequations}
  \begin{align}
    \dot {\hat\vx} & = \hat v_0 \hat \vpsi(\hat t) \\
    \dot{\hat\vpsi}_i & =  \sum_j n_{ij} \hat\vpsi_j^\perp + \hat\vF^\perp
              +  \frac{\chi}{X} \left(\hat\vpsi_i\cdot\nabla\right) \hat\vG_i^\perp  \notag\\
             & \quad + \sqrt{\frac{2T}{J}} \vxi^{\perp}(\hat{t}).
  \end{align}
\end{subequations}
In what follows we shall consider the case of a single particle and a harmonic restoring force $\hat \vF = -k_0 \hat\vx$.  For this force, the overdamped equation (but not the full ISM) reduces to a form that looks like the harmonic oscillator,
\begin{equation}
    \ddot{\hat\vx}  =  - \hat v_0 k_0 \hat\vx^\perp + v_0\sqrt{\frac{2T}{J}} \vxi^{\perp}(\hat t),
    \label{eq:pseudo-HO}
\end{equation}
but with a non-linearity introduced by the projector operator.  Explicitly, going back to dimensionful time,
\begin{equation}
  \eta \ddot\vx = -k_0Jv_0 \vx + k_0Jv_0 (\vpsi\cdot\vx)  \vpsi + \sqrt{2\eta T} \vxi^{\perp}(t).
\end{equation}

\section{Numerical results}

To gain some insight into the behavior of the modified model, we consider the case of a single particle in an external field to study the confinement effects.
We solve the equations of motion numerically in their second-order form \eqref{eq:second-order}.  We have employed an integration scheme used for the ISM in ref.~\cite{cavagna_spatio-temporal_2016}, which has been used in Brownian Dynamics simulations \cite{Rapaport2004} and based on a Velocity Verlet integrator with Lagrange multipliers to enforce fixed speed (see Appendix A for details).  We consider the simplest form of the force to achieve confinement, namely a simple harmonic force
\begin{equation}
  \vF_i = - \frac{1}{v_0} k_0 \vx_i,
\end{equation}
and set $\vG_i=0$.  All results shown are in 2-$d$.

This is the simplest situation one can think of beyond a free particle.  We choose this as the first application of the modified ISM because we can clearly gauge the effects of the position-dependent field, which is the new element in the model, and check that it effectively produces the desired effect, i.e.\ confinement.  But this scenario is also useful for the case of swarms of insects in laboratory conditions, in closed environments (e.g.\ \cite{puckett_searching_2015, ni_intrinsic_2015, gorbonos_long-range_2016, cavagna2023b}) and low densities \cite{reynolds2023}, where interactions can be mostly neglected.  We ask whether inertial effects can be inferred from the trajectory of a single individual, and whether in this situation the ISM can be distinguished from other simple models of a confined particle.  Thus we will compare results at high and very low $\Omega$ (essentially equivalent to comparing the ISM and Vicsek models), and also to two other simple models of confined inertial particles: the Brownian (massive) harmonic oscillator and the active Brownian particle \cite{bechinger2016, zottl2023}.

The Brownian harmonic oscillator (HO) is a useful comparison, partly in view of Eq.~\eqref{eq:pseudo-HO}, but mostly as a null model, being the simplest non-active model for a confined, fluctuating particle.  Clearly the complications of active models are not worth if the resulting phenomenology cannot be distinguished from that of the HO.  The Browinan HO is described by the inertial Langevin equation
\begin{equation}
  m\ddot\vx + \eta \dot\vx + k_0 \vx =\xi(t),
\end{equation}
and the inertial parameter describing the underdamped--overdamped crossover is $\Omega_\text{HO} =k_0 m /\eta^2$.  The dynamic equations for the inertial active Brownian particle (ABP) \cite{enculescu2011} with an harmonic confining field are, in 2-$d$ 
\begin{align}
  m\ddot\vx & = -\eta\dot\vx -k_0\vx + \eta v_0(\cos\theta,\sin\theta) +\xi(t), \\
  \dot\theta &= \zeta,
\end{align}
where $\theta(t)$ is the heading direction, $v_0$ will be held constant and $\xi(t)$ and $\zeta(t)$ are Gaussian white noise processes with zero mean and variances $2T\eta$ and $2D_r$ respectively ($T$ is the temperature and $D_r$ is a rotational diffusion constant).  The inertial parameter is $\Omega_\text{ABP} = k_0m/\eta^2$.

\begin{figure}
  \includegraphics[width=\columnwidth]{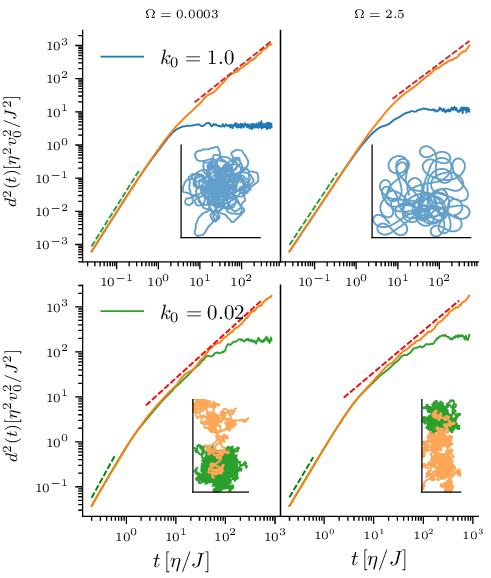}
  \caption{Mean squared displacement $d^2(t)$ vs.\ time for the \ac{ISM} with harmonic confining force at different values of $\Omega$ and $k_0$.  Left panels correspond to $\Omega=3\cdot10^{-5}$, and right panels to $\Omega=2.5$.  The top panels show the MSD for $k_0=1$ (blue) compared to the free particle ($k_0=0$, orange), and the lower panels have $k_0=0.02$ (green) and $k_0=0$.  Note that the orange curves correspond to free particles with different inertia, hence the diffusive regime is reached at different times.  The dashed lines have slope 2 (green) and 1 (orange) and are a guide to identify the ballistic and diffusive regimes.  Time is measured in units of $\eta/J$ and position in units of $v_0\eta/J$.  The MSD is an average over 100 trajectories.  Insets show sample trajectories, with the same color code as the main panels.}
  \label{fig:MSD}
\end{figure}

\begin{figure}
  \includegraphics[width= \columnwidth]{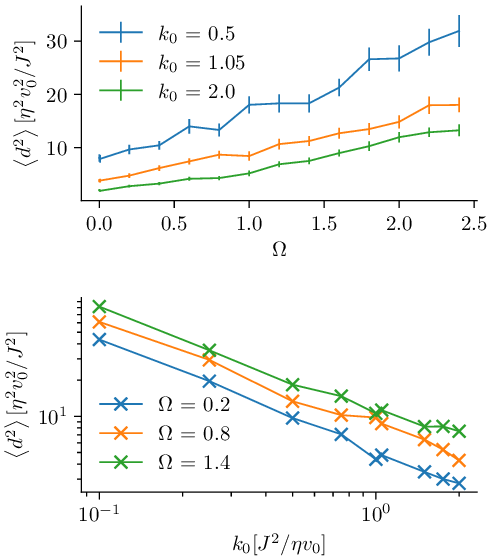}
  \caption{Dependence of the plateau of the mean squared displacement on $\Omega$ and confinement strength $k_0$ for the harmonically confined \ac{ISM}.}
  \label{fig:plateaus}
\end{figure}

A quantity that must be sensitive to confinement effects is the mean squared displacement (MSD) $d^2(t) = [ \vx(t) - \vx(0) ]^2$, which we show in Fig.~\ref{fig:MSD} for the ISM with two different values of the dimensionless parameter $\Omega$ \eqref{eq:defOmega} (recall $\Omega\gg1$ means inertia dominates, while $\Omega\ll1$ corresponds to overdamped systems) and different field strengths (i.e.\ different $k_0$).  The MSD curves are obtained as an average over 100 trajectories, after discarding an initial time so that the MSD is stationary. In all cases, for very short times there is a ballistic regime ($d^2 \sim t^2$) which crosses over to a diffusive ($d^2\sim t$) regime for the free case.  If the confining field is present, the displacement eventually saturates at an $\Omega$-dependent plateau (Fig.~\ref{fig:plateaus}).  For weak enough fields, the diffusive regime can be observed before the plateau, but at high fields the confinement effects manifest before the transition from ballistic to diffusive motion. The presence of the external field brings a qualitative change in the shape of $d^2(t)$ in the expected direction, i.e.\ a plateau appears indicating the particle is confined.  However, there is no qualitative difference between high and low $\Omega$.  The ballistic regime results from the persitence of the velocity (i.e.\ the fact that velocities have a finite correlation time, see below).  In the usual Brownian motion, persistence is a result of an inertial term, but the pure Vicsek particle (the overdamped $\Omega\to0$ limit) also leads to a persistent random walk at low enough temperature, so all the curves of Fig.~\ref{fig:MSD} look qualitatively the same (except for the obvious fact that there is no plateau if $k_0=0$).  Increasing $\Omega$ delays the onset of the diffusive regime, which is hardly observable when both $\Omega$ and $k_0$ are high.  Also, the value of the plateau itself depends on $\Omega$ and on $k_0$ (Fig.~\ref{fig:plateaus}).  The asymptotic $d^2$ goes as $\sim 1/k_0$, similarly to the Brownian harmonic oscillator where by equipartition one has $\langle x^2\rangle = k_B T/k_0$.  On the other hand, the plateau grows (almost linearly in the explored range) with $\Omega$, quite differently from the Browinan oscillator, where the confinement scale is independent of the mass.  The ABP (not shown) behaves very similar to the ISM (the plateau goes as $\sim 1/k_0$ and depends on $\Omega_\text{ABP}$), except that the plateau dependence with inertia is clearly superlinear for $\Omega_\text{ABP}$ greater than approximately 2.

\begin{figure}
  \includegraphics[width=\columnwidth]{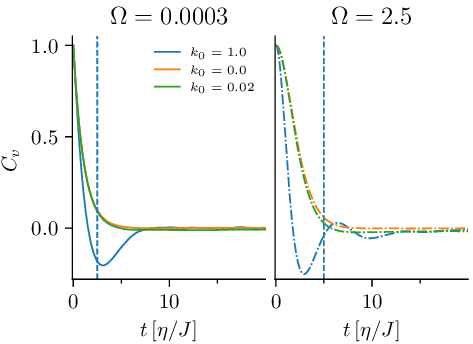}
  \caption{Velocity time correlation function $C_v(t)$ for $\Omega=3\cdot10^{-4}$ (left) and $\Omega=2.5$ (right) and three values of $k_0$.  The vertical dotted line indicates the time at which the system with $k_0=1$ starts feeling the confinement effects of the field (Fig.~\ref{fig:MSD}).  For the lower value of $k_0$, confinement is only observed around $t\approx 20$ to 25$\,\eta/J$, which is out of the scale of the abscissa.}
  \label{fig:vcorr}
\end{figure}

\begin{figure}
  \includegraphics[width=\columnwidth]{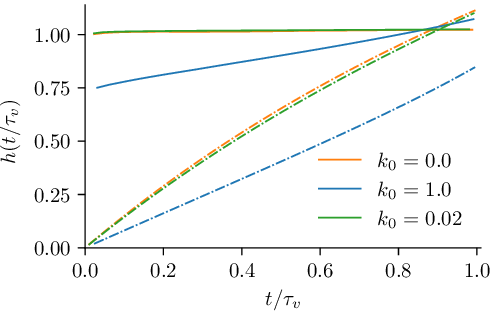}
  \caption{The function $h(x)=-\log C_v(x)/x|_{x=t/\tau}$ for $\Omega=3\cdot10^{-4}$ (full lines) and $\Omega=2.5$ (dotted lines).  Colors indicate values of $k_0$ as in Fig.~\ref{fig:vcorr}.}
  \label{fig:h}
\end{figure}

\begin{figure}
  \includegraphics[width=\columnwidth]{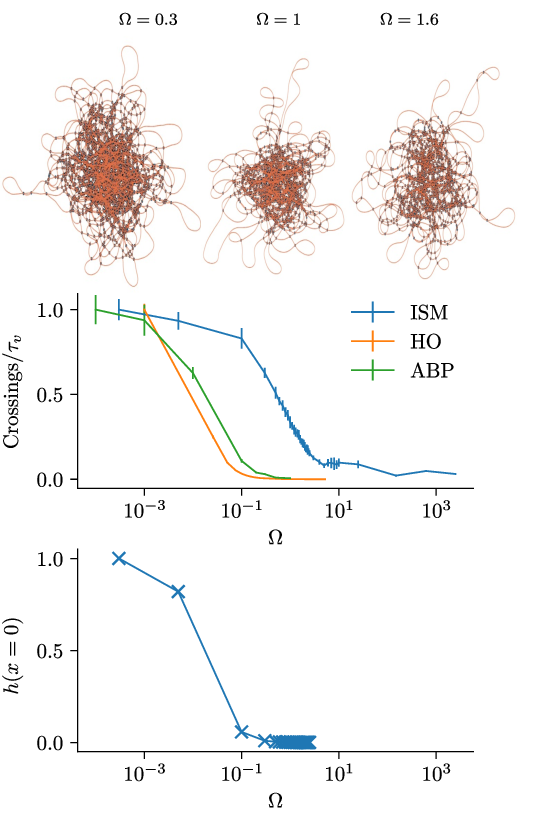}
  \caption{Number of times a trajectory crosses itself during one velocity correlation time vs.\ inertial parameter for the ISM, harmonic oscillator and ABP (middle panel).  The values have been normalized to 1 at $\Omega=0$ to be able to plot them together, since the value at small $\Omega$ varies much for the three models.  Bottom panel: $h(x)$ (see Fig.~\ref{fig:h}) extrapolated to $x=0$ for the ISM.  Top panel: example trajectories near the crossover in crossings.}
  \label{fig:crossings}
\end{figure}

The effects of confinement can also be noticeable in the velocity correlation function $C_v(t) = \langle \vv(t)\cdot\vv(0) \rangle$ (Fig.~\ref{fig:vcorr}).  If the confining field strength is large enough, there is a range of times for which anticorrelation is observed and followed, at high $\Omega$, by damped oscillations.  The first anti-correlation minimum should not be mistaken for an inertial effect: it is due to the fact that the velocity changes direction due to the confinement force, and it is actually observed both at low and high $\Omega$ for large enough $k_0$.  For $k_0=1$ it can be seen that $C_v(t)$ has its first minimum around $t\approx 3$, which is the time at which confinement effects start to be noticeable in $d^2(t)$ (cf.\ top panels of Fig.~\ref{fig:MSD}).  At this time, the diffusive regime has not yet been reached, and the correlation function of the free particle is still non-vanishing.  In contrast, for weaker fields (e.g.\ $k_0=0.02$ in Fig.~\ref{fig:vcorr}) it can happen that the MSD starts deviating towards the plateau already in the diffusive regime, when the velocity has lost correlation (as measured by $C_v(t)$ for the free case).  In this case no anticorrelation minimum is observed.  Inertial effects manifest, for large $k_0$, in the presence of damped oscillations in $C_v(t)$, lasting roughly during the transition from the ballistic regime to the plateau in the MSD.  For all $k_0$, when $\Omega$ is large the correlation function has a flat derivative as $t\to0$, a sign of second-order dynamics (see discussion in \cite{cavagna_dynamic_2017}).  This can be seen plotting $h(x) = - \log C(x) /x$ with $x=t/\tau$ and $\tau$ is the correlation time (Fig.~\ref{fig:h}).  $h(x)$ tends to a constant for $x\to0$ if $C(t)$ resembles a simple exponential for short times (overdamped system), or to 0 if $C(t)$ has a flat derivative.  For the correlation time $\tau$ we have used the spectral definition \cite{halperin_scaling_1969},
\begin{equation}
   \int_0^\infty \!\! \frac{dt}{t} \, \frac{C(t)}{C(t=0)}
   \sin\left(\frac{t}{\tau}\right) = \frac{\pi}{4}.
\end{equation}
The extrapolation of $h(x)$ to $x=0$ is appreciably different from 0 only for rather low values of $\Omega$, dropping rather sharply to very low values around $\Omega=10^{-1}$ (see Fig.~\ref{fig:crossings}).

\begin{figure}
  \includegraphics[width = \columnwidth]{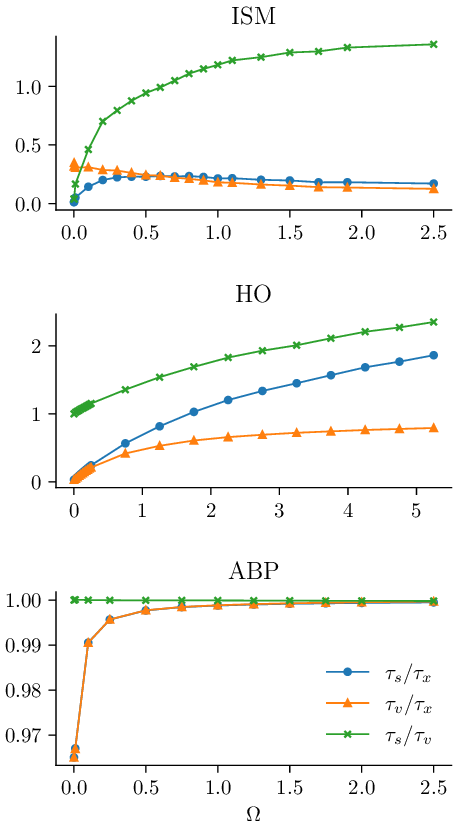}
  \caption{Correlation times ratios vs.\ inertial parameter for the ISM (top), 2-$d$ stochastic harmonic oscillator (middle) and active Browinian particle (bottom).}
  \label{fig:taus}
\end{figure}

The velocity time correlations look similar to those of an harmonic oscillator with Langevin dynamics.  However, the behavior of the active system is different from the harmonic oscillator's.  One way to see this is considering the correlation times of the three dynamical variables position $\vx$, velocity $\vv$ and spin $\vs$.  Since the spin can be computed from the trajectories as $\vs = \mu \vv\times \dot\vv$, this definition can also be applied to trajectories of the harmonic oscillator and the ABP, even if the spin does not play a significant role in those models.  So we can compute the time correlation functions for position $C_x(t)$, velocity $C_v(t)$ and spin $C_s(t)$ in the three systems.  The ratios $\tau_v/\tau_x$ and $\tau_s/\tau_x$ show quite different behavior as a function of $\Omega$, $\Omega_\text{HO}$ or $\Omega_\text{ABP}$  (Fig.~\ref{fig:taus}).  In the HO, both ratios increase monotonically as the system becomes more underdamped, while in the ISM $\tau_v/\tau_x$ is monotonically decreasing and $\tau_s/\tau_x$ reaches a maximum near $\Omega=1$ and then decreases.  The third ratio, $\tau_s/\tau_v$, is monotonically increasing in both cases, but as $\Omega\to0$ it tends to 1 for the HO while it goes to zero for the ISM, as in the latter the spin correlation time vanishes more quickly than that of the velocity. In the ABP, $\tau_s$ and $\tau_v$ are always very similar and are both smaller than $\tau_x$.  Their value relative to $\tau_x$ increases monotonically with $\Omega_\text{ABP}$, and as $\Omega_\text{ABP}$ grows all correlation times tend to coincide.

If the relaxation times could be computed at several values of the inertia, it would be possible to discriminate among the three models.  However, this is harder when working at fixed $\Omega$ (which would be the typical situation in experimental observations).  But in principle one could use $h(x)$ for the velocity correlation function $C_v(t)$ to estimate whether $\Omega$ is high or low, then study the relaxation time ratios.  If $h(x)$ stays finite for $x\to0$ (low $\Omega$), then $\tau_s/\tau_v<1$ is not compatible with simple harmonic motion, and neither is the presence of a ballistic regime in $d^2(t)$.  At high inertia instead, $\tau_s/\tau_x$ and $\tau_v/\tau_x$ both less than one and of similar value would be incompatible with an HO.  Also $\tau_s/\tau_v$ much different from 1 would exclude the ABP, at any value of inertia.

Finally, we have considered a measure of the trajectories' shape.  The sample trajectories in Fig.~\ref{fig:MSD} show that in the inertial case the particle finds it more difficult to turn back on itself, and responds to confinement making turns with smoother curvature than in the over-damped case.  This suggests that measuring the number of times the trajectory intersects with itself might be a way to detect inertial effects directly from the trajectory.  We show the number of self-intersections during one velocity correlation time in Fig.~\ref{fig:crossings}.  It is clear that at low inertia the trajectory crosses itself much more often than at high $\Omega$.  This tendency is also observed in the harmonic oscillator and the ABP. A rather sharp drop is observed at a model-dependent value of $\Omega$.  Inspecting the trajectories directly, it is clear that this number decreases as trajectories turn less on themselves and tend to make smoother loops rather than sharp turns.  However, it is not evident that there is a clear qualitative change in their shapes across the crossover.  Fig.~\ref{fig:crossings} also shows the value $h_0$ of $h(x)$ extrapolated to $x=0$.  Although it also displays a rather sharp crossover, it occurs at a value of $\Omega$ about a hundred times smaller than that of the crossings.  The two quantities thus reflect different properties of the trajectories.

\section{Discussion and conclusions}

We have shown how the \acl{ISM} equations of motion must be modified to include position-dependent forces.  The most compact formulation is in terms of velocity and spin, Eqs.~\eqref{eq:ISM-field-final}, where the positional forces enter only in the equation for the spin. The ISM was proposed as an extension of the Vicsek model, and thus it introduces activity in the same way, which is perhaps the simplest one can imagine: by imposing a hard constraint on the particles' speed.  This, plus the ISM's own Hamiltonian structure, greatly limits the way in which position-dependent forces may be added to this model.  Another approach to introducing activity is to endow the particles with a ``heading'' or ``spin'', which is an internal degree of freedom related to but different from the velocity, as in active Brownian particles \cite{bechinger2016, zottl2023}.  This is also the approach of the purely Hamiltonian, conservative, active model of \cite{bore2016}.  Interestingly, in that model the Hamiltonian structure also places limits in the kind of interaction term one can consider (in that case the interaction between velocity and spin \cite{casiulis2020a}).

As a simple application of the new equations, we have considered a single harmonically confined active particle, and shown how inertia can be detected in this simple case without collective effects.  We have shown the confined ISM shows behavior qualitatively different from both the Brownian harmonic oscillator (our null, non-active, model for confined motion) and the active Brownian particle, another simple model for self-propelled particles.  The three models show a plateau in the mean-square displacement that depends in the same way on the strength of the confining field (measured by the value of the harmonic constant $k_0$).  However, the active models distinguish themselves from the harmonic oscillator in that the plateau also depends on the value of the inertial parameter $\Omega$ or $\Omega_\text{ABP}$.  Up to the explored values of $\Omega$, this dependence is linear for the ISM, while a non-linear regime is evident in the ABP.

We have also studied the time correlation functions that can be computed from the trajectory.  The effect of strong confinement is the appearance of an anti-correlation minimum in the velocity temporal self-correlation. This anti-correlation appears when the velocity is forced to change direction at times shorter than the correlation time of the free particle under otherwise identical parameters; it appears in over- and under-damped systems and is unrelated to inertia.  When present, inertia manifests in the slope of the correlation function at short times (Fig.~\ref{fig:h}), and in the presence of damped oscillations beyond the first anti-correlation minimum. Such damped oscillations were recently reported for male \emph{Anopheles gambiae} (malaria mosquitoes) in laboratory swarms \cite{cavagna2023b}.  Also, the trajectories recorded in that work appear to show relatively smooth loops, similar to the simulated trajectories at high $\Omega$ (Fig.~\ref{fig:MSD}), rather than the more tortuous trajectories of the low-$\Omega$, Vicsek-like case.  Thus, inertia seems an important ingredient to include in the modelling, even for the study of single individuals, making the present developments useful for attempting to apply the \ac{ISM} to the analysis of these and other experiments on laboratory-confined swarms (e.g.\ \cite{ni_intrinsic_2015, gorbonos_long-range_2016}).  However, the velocity correlation function alone is not enough to distinguish among the harmonic oscillator, the ISM and the ABP, as the different parameters in each model can be tuned to give very similar correlation functions.  We have shown that the models display clear differences when one examines the \emph{trends} of the relaxation times and the plateau in the mean squared displacement (Figs.~\ref{fig:plateaus} and~\ref{fig:taus}).   Unfortunately in the experimental case one will usually not be able to tune $\Omega$, so discriminating among models will be harder, but some conclusions can be drawn examining together the MSD and the three relaxation times.

As a next step, it is clearly of interest to study systems of many particles with open boundary conditions, maintaining a finite density through confinement or cohesive interactions.  This can be done with the present formulation of the \acl{ISM}, and it is worth pursuing to achieve a better description of experimental results as well as to gain further theoretical understanding of the collective properties of active models.  As an example, in a series of recent papers \cite{gonzalez-albaladejo2023, gonzalez-albaladejo2023a, gonzalez-albaladejo2023b}, it has been claimed that, for the Vicsek model, replacing periodic boundaries with an harmonic confinement alters the behavior of the model near ordering, giving rise to a phase transition characterized by scale-free chaos and an extended criticality region and yielding different static and dynamic critical exponents.  Open conditions thus deserve deeper inquiry, both for over- and under-damped inertial systems.  Finally, position-dependent forces can also be used to investigate the response of a swarm to external perturbations that do not directly alter the velocities.

In summary, we have shown how positional forces should be added to the Inertial Spin Model, preserving its Hamiltonian structure, and presented evidence that a single confined particle displays qualitatively different behavior in the ISM than in two other simple models of confinement (one equilibrium, one active).  This development makes it possible to use the ISM without periodic boundary conditions, by introducing either a confining field or cohesion.  The addition of these new ingredients allows to apply the ISM to a broader range of conditions, and to simulate more closely various aspects of biological groups.

\acknowledgements

We thank A.~Cavagna, I.~Giardina, S.~Melillo and M.~L.~Rubio Puzzo for helpful discussions.  This work was supported by Agencia I+D+i (Argentina) PICT2020/00520, CONICET (Argentina) PIP2022/11220210100731CO and Universidad Nacional de La Plata (Argentina) UNLP 11/X787.  SC was supported in part by Comisi\'on de Investigaciones Cient\'\i{}ficas, Provincia de Buenos Aires (Argentina) and GP was partially supported by ERC Advanced Grant RG.BIO (785932).

\appendix

\section{Details of the numerical simulations}

To simulate the ISM or the Brownian harmonic oscillator we integrate numerically the corresponding differential equations. In the case of the ISM model, since we have the constraint $\mathbf{v}^2(t)=v_0$ we use an integration scheme used in Brownian Dynamics which allows for an exact implementation of said constraint via Lagrange multipliers, and which in the underdamped case $\Omega\to\infty$ reduces to the velocity Verlet integrator widely used in Molecular Dynamics due to its good energy conservation properties and computational affordability \cite{Rapaport2004}. The only drawbacks is that the overdamped case with $\chi$ strictly equal to 0 cannot be integrated using this method.  This scheme has been employed before in simulations of the ISM, e.g.\ in Ref.~\cite{cavagna_spatio-temporal_2016}.

To arrive at an integration algorithm we start from the second order equation \eqref{eq:second-order}, which we rewrite as
\begin{equation}
    \frac{d^2\mathbf{v}}{dt^2} = \frac{v_0^2}{\chi} \left[ \mathbf{F} + \mathbf{F}_{v} + \mathbf{f}_c \right]
    \label{eq:apppendix-differentialEq}
\end{equation}
where the first term represents the position-dependent external (in our particular simulation case $\mathbf{F} = -\frac{k_0}{v_0}\mathbf{x} $), the next term includes the random and viscous forces and $\mathbf{f}_c$ is the constraint force.  At the continuous level it is given by the rest of \eqref{eq:second-order}, but for the numerical integration it is computed as explained below, so that the constraint is enforced exactly.  To obtain the discretized equations we integrate \eqref{eq:apppendix-differentialEq} assuming $F$ varies linearly with time in a small interval $\Delta t$.  The term $\mathbf{f}_c$ is disregarded at first and later reintroduced  as explained below. Defining $\mathbf{a} = \frac{d\mathbf{v}}{dt}$, $\mathbf{b} = \frac{d \mathbf{a}}{dt}$ one arrives at
\begin{align}
&\mathbf{r}(t+\Delta t) = \mathbf{r}(t)+\Delta t \mathbf{v}(t) \\
\begin{split} 
&\mathbf{v}(t+\Delta t) = \mathbf{v}(t)+\Delta t c_1\mathbf{a}(t)+(\Delta t)^2 c_2 \mathbf{b}(t) \\
&+ (\Delta t)^2 c_2 \zeta_1\mathbf{v}(t) + \Theta_v(t) 
\end{split}   \\ 
\begin{split}
&\mathbf{a}(t+\Delta t) = c_0 \mathbf{a}(t)+(c_1-c_2)\Delta t \left[ \mathbf{b}(t)+\zeta_1 \mathbf{v}(t) \right] + 
\\ &c_2 \Delta t \left[ \mathbf{b}(t+\Delta t) + \zeta_2 \mathbf{v}(t+\Delta t)\right] + \Theta_a(t)
\end{split}  \\
&\mathbf{b}(t+\Delta t) = \frac{v_0^2}{\chi} \mathbf{F}(\mathbf{r}(t+\Delta t))
\end{align}
where $\lambda$ and $\mu$ are related to the constraint, the other constants result from the integration as
\begin{align}
    &c_0 = e^{ - \eta \Delta t/ \chi} \\
    &c_1 = \frac{\chi }{\eta \Delta t} (1-c_0)\\
    &c_2 = \frac{\chi }{\eta \Delta t}  (1-c_1),
\end{align}
and $\Theta_v$ and $\Theta_a$ are random variables related to the random force. They are independent for each axis and each pair of components is drawn form a bivariate Gaussian distribution with zero first moments and second moments given by
\begin{align}
    \left<\Theta_v^2 \right> &= \frac{T \chi}{ \eta^2} \left( 2\frac{\eta }{\chi}\Delta t-3+4c_0-c_0^2 \right) \\
    \left< \Theta_a^2 \right> &= \frac{T}{ \chi} \left( 1-c_0^2 \right) \\
    \left< \Theta_v \Theta_a \right> &= \frac{T}{\eta} \left(1-c_0 \right). 
\end{align}
The discrete equations reduce to the velocity Verlet integrator for Molecular Dynamics \cite{allen1987computer,swope1982computer} in the underdamped $\eta \rightarrow 0$ limit.

The constraint is enforced as in the RATTLE algorithm \cite{andersen1983rattle}, only that since the constraints on each particle are independent, the Lagrange multipliers can be found analytically and there is no need of an iterative procedure. Imposing $\mathbf{v}^2(t+\Delta t) = v_0^2$ and $\mathbf{v}(t+\Delta t) \cdot \mathbf{a}(t+\Delta t) = 0$ one obtains
\begin{align}
    \zeta_1 &= \frac{\omega_+-1}{(\Delta t)^2c_2}, \\
    \zeta_2 &= - \frac{\mathbf{v}(t+\Delta t)\cdot \mathbf{a}'(t+\Delta t)}{c_2v_0^2\Delta t},
\end{align}
where $\omega_+$ is the positive root of
\begin{align}
    &v_0^2\omega^2+2\mathbf{v}\cdot \Delta \mathbf{v} \omega + \Delta v^2 = v_0^2, \\
    &\Delta \mathbf{v} = c_1 \Delta t \mathbf{a}(t)+c_2 (\Delta t)^2\mathbf{b}(t),
\end{align}
and $\mathbf{a}'(t+\Delta t)$ is equal to $\mathbf{a}(t+\Delta t)$ without the term proportional to $\zeta_2$. 

Each step is performed in two stages, as in the velocity Verlet scheme \cite{allen1987computer}. First the random variables are drawn, $\mathbf{r}$ is updated, $\mathbf{a}$ is partially updated using only the terms that depend on quantities evaluated at $t$, $\mathbf{v} $ is updated and then the constraint terms are computed and applied. Second the forces for the new positions and  velocities are computed, and finally the update of $\mathbf{a}$ is completed.

For the results reported here, we have set $\Delta t = 0.001$ and the friction coefficient and temperature respectively to $\eta = 1$ and $T=1$.

\bibliography{refs}

\end{document}